\documentclass[aps,prl,reprint,superscriptaddress]{revtex4-2}
\usepackage{graphicx}
\usepackage[dvipsnames]{xcolor}
\usepackage[colorlinks=true,citecolor=blue]{hyperref}

\begin{document}

\title{Hot Carrier Nanowire Transistors at the Ballistic Limit}

\author{M. Kumar}
\author{A. Nowzari}
\affiliation{NanoLund and Solid State Physics, Lund University, Lund, Sweden}

\author{A. R. Persson}
\affiliation{NanoLund and Centre for Analysis and Synthesis, Lund University, Sweden}

\author{S. Jeppesen}
\affiliation{NanoLund and Solid State Physics, Lund University, Lund, Sweden}

\author{A. Wacker}
\affiliation{NanoLund and Mathematical Physics, Lund University, Lund, Sweden}

\author{G. Bastard}
\affiliation{Physics Department ENS-PSL, 24 rue Lhomond F75005, France}

\author{R. Wallenberg}
\affiliation{NanoLund and Centre for Analysis and Synthesis, Lund University, Sweden}

\author{F. Capasso}
\affiliation{John A. Paulson School of Engineering and Applied Sciences, Harvard University, United States}

\author{V. F. Maisi}
\affiliation{NanoLund and Solid State Physics, Lund University, Lund, Sweden}

\author{L. Samuelson}
\affiliation{NanoLund and Solid State Physics, Lund University, Lund, Sweden}
\affiliation{Institute of Nanoscience and Applications, Southern University of Science and Technology, Shenzhen, China}

\date{\today}

\begin{abstract}
We demonstrate experimentally non-equilibrium transport in unipolar quasi-1D hot electron devices reaching ballistic limit. The devices are realized with heterostructure engineering in nanowires to obtain dopant- and dislocation-free 1D-epitaxy and flexible bandgap engineering. We show experimentally the control of hot electron injection with a graded conduction band profile and subsequent filtering of hot and relaxed electrons with rectangular energy barriers. The number of electron passing the barrier depends exponentially on the transport length with a mean free path of $200 - 260$~nm and reaches ballistic transport regime for the shortest devices with $70\ \%$ of the electrons flying freely through the base electrode and the barrier reflections limiting the transport to the collector.
\end{abstract}

\maketitle

Hot carriers open up many interesting device concept paradigms based on their ballistic transport in semiconductors~\cite{barati2017, brongersma2015}. In the ballistic transport, charge carriers are free to move in a host material without losing their energy. Thus, for instance high-speed operation is anticipated, as scattering does not limit the carrier flow. To obtain hot electron devices, band-gap energy of a semiconductor needs modulations at energies much larger than the thermal energy $kT$, to generate and detect highly non-equilibrium electrons. Typically band-gap engineering has been done by either gating~\cite{hohls2006,rossler2014,shur1979}, doping~\cite{hayes1985,levi1985} or using Schottky or tunnel barriers in metal-oxide-semiconductor structures~\cite{heiblum1981,capasso1990,brill1996}. While being successful in making ballistic devices, the above-mentioned approaches come with major constrains. For doped structures, the dopants act as scattering centers. Additional scattering and energy relaxation therefore limit the free flight of the carriers and is a serious constrain for the ballisticity. Doping also typically varies smoothly in the structure prohibiting to define sharp boundaries between different operational parts of the devices. On the other hand, the Schottky barrier based devices generate hot electrons via tunnelling process. This together with the inability to adjust the barrier height and thickness hinder to choose and set the hot carriers and their detector barrier to specific energies. In this letter, we use 1D epitaxy in semiconductor nanowires for the band-gap engineering. The 1D epitaxy allows us to tailor the band-gap by combining different materials into heterostructures without limitations arising from strain-induced dislocations or dopants~\cite{bjork2002,wu2002,gudiksen2002,bjork2002b,thelander2003,nylund2016}. We make devices with the 1D epitaxy where up to $70\ \%$ of the electrons fly ballistically at room temperature and determine the mean free part and the reflection probability for the electrons at the barrier. The findings are consistent with theoretical predictions.

\begin{figure}[h!t]
    \centering
    \includegraphics[width=0.42\textwidth]{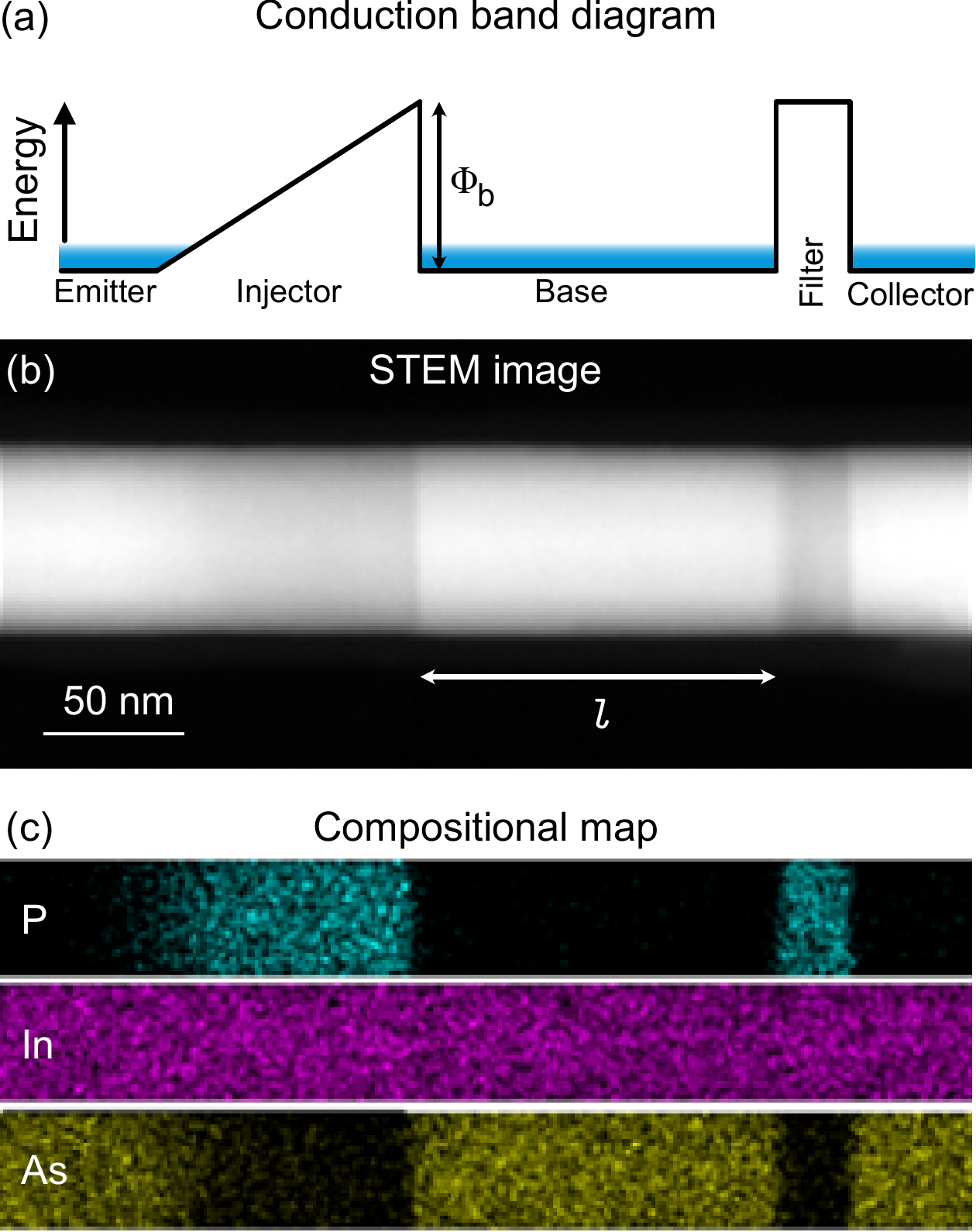}
    \caption{(a) The conduction band diagram of the ballistic hot electron system consisting of a graded potential barrier as a hot electron injector and a rectangular barrier for filtering the electrons. (b) Scanning transmission electron microscope image of a nanowire heterostructure obtained through growth engineering showing $\mathrm{InAs}_{1-x}\mathrm{P}_x$ based quasi-1D hot electron injector and InP based energy filter. (c) Compositional map of In, As and P along the nanowire.}
    \label{fig:bandstruct}
\end{figure}

\begin{figure}[t]
    \centering
    \includegraphics[width=0.42\textwidth]{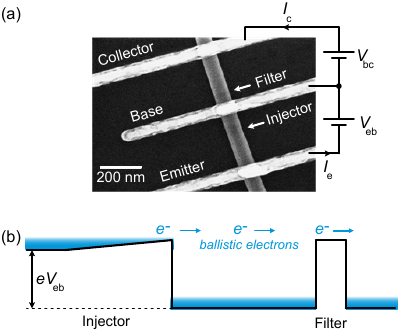}
    \caption{(a), A scanning electron micrograph showing one of the measured unipolar 3-terminal hot electron nanowire devices. The middle base electrode needs to be positioned and defined sharply between the injector and the energy filter spaced by a distance $l$. Voltage $V_\mathrm{eb}$ biases the injector and $V_\mathrm{bc}$ the filter. (b) Band diagram under typical operation. The voltage $V_\mathrm{eb}$ lifts the energy of the emitter electrons and evens the graded injector giving rise to electron injection. Ballistic electrons fly over the filter to an unbiased ($V_\mathrm{bc} = 0$) collector.}
    \label{fig:dev}
\end{figure}

The conduction band profile of our hot electron injector-collector system is presented in Fig.~\ref{fig:bandstruct}a. A graded potential barrier on the left forms a hot electron injector. Electrons will be injected close to the maximum value of the barrier by raising the chemical potential of the graded side, see Fig.~\ref{fig:dev}b, which evens out the grading and allows electrons to flow through the graded segment. As a detector, we use a rectangular barrier that has the same height as the injector so that ideally only ballistic electrons at high energy pass. Electrons that relax in the middle region will be blocked by the rectangular barrier which we grow thick enough (thickness $20$ nm with barrier height of approximately $500$ meV in InAs/InP heterointerface) to suppress tunnelling at low energies. Thus, the electrical current after the filtering barrier measures the number of ballistic electrons that fly over the barrier. The sharp interfaces also define the length $l$ of the base electrode between the injector and filter unambiguously.

For realizing the band-diagram of Fig.~\ref{fig:bandstruct}a, we grow InAs/InAsP nanowire heterostructures with chemical beam epitaxy (CBE) allowing for the demanding abrupt hetero-interfaces~\cite{fuhrer2007} and smooth graded barriers~\cite{nylund2016} within the same growth run. The InAs and InP have additionally a large conduction band-offset, low effective mass and high electron mobility which are preferred properties for the hot electron devices~\cite{persson2006}. The growth took place on InAs(111)B substrates with deposited Au aerosol particles catalysing the nanowire growth~\cite{ohlsson2001} and defining the diameter and length of the order of several tens of nanometers and few microns, respectively (see supporting information containing a SEM image for typical as-grown nanowires heterostructures). The graded injectors in InAs based nanowires were formed with graded $\mathrm{InAs}_{1-x}\mathrm{P}_x$ through in-situ modulation of chemical composition $x$ along the length of nanowires. By adjusting the flow rate for trimethylindium (TMIn) as group-III source, pre-cracked tertiarybutylarsine (TBAs) and pre-cracked tertiarybutylphosphine (TBP) as Group-V sources, the ternary composition formed. To limit tunnelling through the sharp triangular top of the injector barrier the highest phosphorous content based ternary was extended at fixed composition by about a few $10$'s of nm before a sharp termination of the barrier. After growing the subsequent InAs segment defining the flight length $l$ for the electrons, the rectangular energy filter barrier was grown with InAsP heterostructure segment in a nanowire. The growth conditions required careful tuning to provide the right ternary composition, sharp hetero-interfaces, correct dimensions, and to maintain the entire heterostructure in a straight axial morphology. Figures~\ref{fig:bandstruct}b and \ref{fig:bandstruct}c show a scanning transmission electron micrograph, and compositional analysis of the structure that highlight clearly the key features and dimensions. To estimate the barrier height for both the injector and filter, we estimate the uppermost phosphorus content to be around $x \approx 0.8$ in both segments corresponding to conduction band offset of about $0.5$ eV~\cite{nylund2016,persson2006}.

After the growth, we transferred the structures to a $\mathrm{Si/SiO_2}$ chip and made ohmic contacts. We made three contacts: an emitter contact before the injector, a collector contact after the rectangular barrier and a common base contact in between. These form a transistor configuration as presented in Fig.~\ref{fig:dev}. We used electron beam lithography to define and position the contacts allowing to contact devices with base lengths down to $l = 80$ nm. Before the deposition of the 25 nm/ 125 nm Ni/Au contacts with thermal evaporation, sulphur passivation~\cite{suyatin2007} was used for obtaining low contact resistance of the order of $100\ \mathrm {\Omega}$ measured from similar contacts made to the plain InAs segment. 

The generation of ballistic electrons in the structure takes place by applying a bias voltage $V_\mathrm{eb}$ to the graded barrier. This voltage lifts the electrons on the left of the the injector to higher energies and evens out the graded barrier as shown in Fig.~\ref{fig:dev}b. When the energy from the bias voltage $eV_\mathrm{eb}$ exceeds the barrier height $\Phi_b$, the injector barrier no longer limits the current flow leading to high injection current with electrons at energy $\Phi_b$ in the base region. We measure the injection current $I_e$ from the emitter side. The energetic electrons have two possible scenarios at the base. Either they continue at high energy over the filtering barrier contributing to collector current $I_c$ or they relax at the base regime, get trapped there and flow away to ground from the base contact.

\begin{figure}[t]
    \centering
    \includegraphics[width=0.48\textwidth]{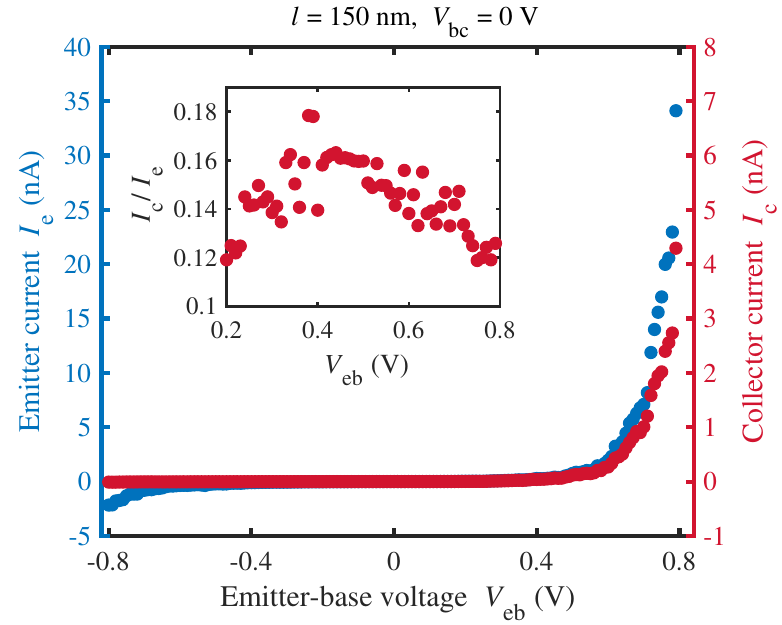}
    \caption{Measured current-voltage curves for a device with base length $l = 150$ nm. The emitter current $I_e$ and collector current $I_c$ were measured simultaneously as a function of the injector voltage $V_\mathrm{eb}$. The energy filter was kept unbiased, $V_\mathrm{bc} = 0$. The inset shows transfer ratio $I_c /I_e$ at the injection regime. All measurements took place at room temperature.}
    \label{fig:IV}
\end{figure}

\begin{figure}[t]
    \centering
    \includegraphics[width=0.48\textwidth]{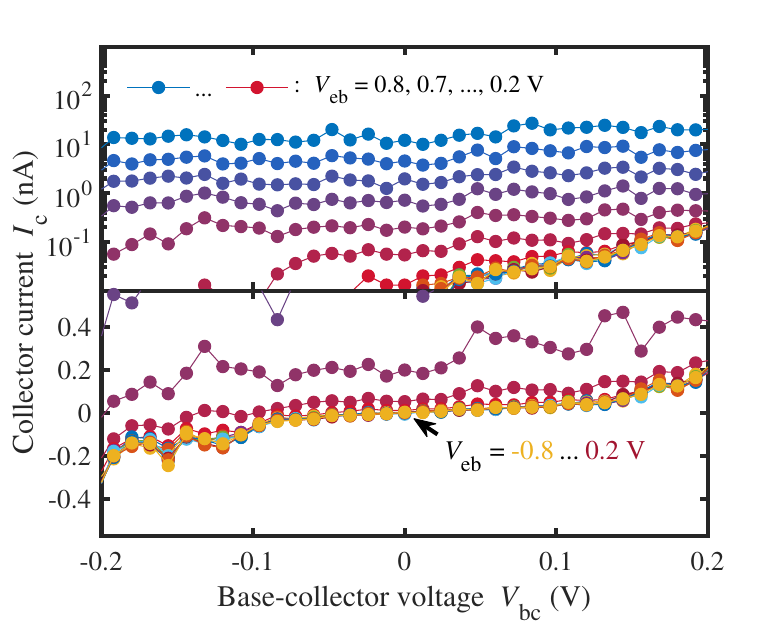}
    \caption{The collector current $I_c$ as a function of the base-collector voltage $V_\mathrm{bc}$ for emitter-base voltages $V_\mathrm{eb} = -0.8 ... 0.8$~V. The device has a base lenght of $l = 160$~nm.}
    \label{fig:Vbc}
\end{figure}

Figure~\ref{fig:IV} presents transport data for the device of Fig.~\ref{fig:dev} with $l = 150$~nm. We indeed see vanishing injector current $I_e$ at low bias voltage and at $V_\mathrm{eb} > 0.5$~V, the current increases steeply, consistent with the estimated barrier height of $\Phi_b \approx 500$~meV. These findings and numbers are consistent with our earlier study of the graded barrier as electrical diodes~\cite{nylund2016}. On the collector side, the current $I_c$ stays also vanishingly small at low $V_\mathrm{eb}$ and increases proportional to the emitter current $I_e$. The inset presents the proportionality as a transfer ratio $I_c/I_e$. We observe that the proportionality stays within $I_c/I_e = 14\ \% \pm 2\ \%$ over the whole injection range. Interestingly, if we apply a large reverse bias $V_\mathrm{eb} < -0.6$~V to the emitter, a small leakage current $I_e$ appears at the emitter but no current at the collector side. This reverse biasing removes equilibrium electrons at low energies from the base and does not create high energy ones. As the collector current remains at $I_c = 0$, the collector side does not respond to these low energy excitations. As a further proof of the energy selectivity, Fig.~\ref{fig:Vbc} shows the collector current $I_c$ dependence on the base collector voltage $V_\mathrm{bc}$ with and without injection. Without injection at $V_\mathrm{eb} < 0.2$~V, we observe a  diminishing collector current $I_c$ below $0.02$~nA level up to $V_\mathrm{bc} = \pm 50$~mV. At larger $V_\mathrm{bc}$, an exponentially increasing leakage current $I_c$ through the energy filter arises. By using $V_\mathrm{bc} = 0$ for the injection experiments, we minimize these leakage contributions. With emission, the collector current $I_c$ depends only weakly on the collector bias voltage $V_\mathrm{bc}$: a bias voltage variation of $V_\mathrm{bc} = \pm 50$~mV  changes the collector current by $20\ \%$ or less in the high emission current regime.

\begin{figure}[t]
    \centering
    \includegraphics[width=0.48\textwidth]{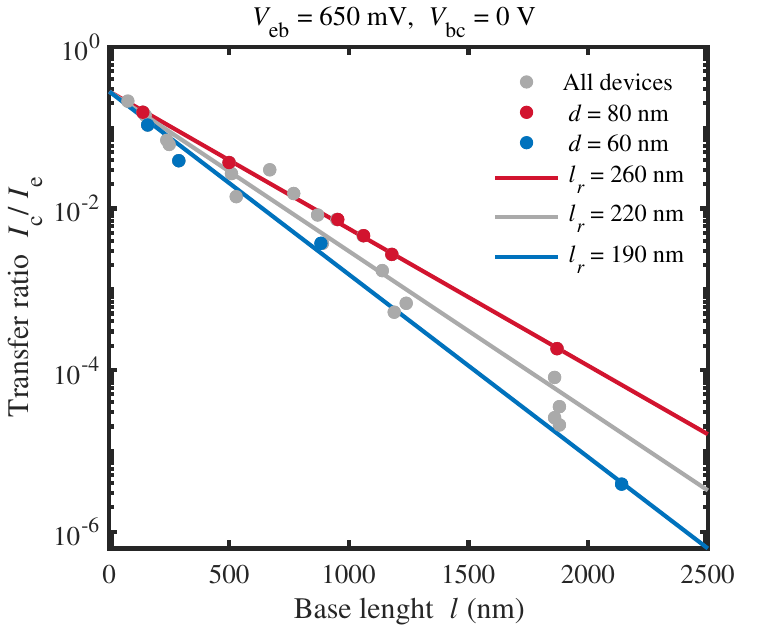}
    \caption{Base length $l$ dependence of the transfer ratio $I_c/I_e$. Dots are experimental data from several devices and the solid lines are fits to exponential dependence $I_c/I_e = T \exp(-l/l_r)$. The red and blue colored data shows the results with nanowires sorted to two diameters $d = 60 ... 65$ nm and $76 ... 82$ nm respectively.}
    \label{fig:ldep}
\end{figure}

We now turn into investigating the base length $l$ dependence in order to study the ballistic transport characteristics through the base and over the rectangular barrier. For that, we repeated the device fabrication and measurements for varying base lengths. Figure~\ref{fig:ldep} summarizes the findings. We observed consistently a collector current that was proportional to the injection current and depended only weakly on the other parameter values as above. However, the transfer ratio $I_c/I_e$ depends strongly on the base length $l$. We see from Fig.~\ref{fig:ldep} that the measured $l = 80\ –\ 2000$ nm range results in more than four orders of magnitudes change to $I_c/I_e$ and scales exponentially as $I_c/I_e = T \exp(-l/l_r)$ as shown by the solid lines fitted to the data. From the fit, we determine energy relaxation length $l_r = 220$ nm and transmission probability $T = 0.28$ at the filter for ballistic electrons. For the shortest devices with $l = 80$ nm, we obtain $\exp(-l/l_r) = \exp(-80\ \mathrm{nm} / 220\ \mathrm{nm}) = 70\ \%$ of the electrons ballistically flying to the filter and the transmissivity of $T = 0.28$ sets predominantly the transfer ratio $I_c/I_e = 0.19$. The observed transmission probability $T$ is consistent with theoretical values: with a rectangular barrier ($\Phi_b = 500$~meV, thickness $a = 20$~nm) a thermal electron distribution above the barrier provides an avergare transmission of $T = 0.3$ in good agreement with the measured value~\cite{davies2006}, see methods.

With the red and blue points in Fig.~\ref{fig:ldep}, we sorted the devices into two categories with different diameters $d$. We observe that majority of the scattering of the data is explained by diameter dependence with smaller diameter $d = 60$~nm leading to shorter mean-free path of $l_r = 190$~nm compared to larger diameter $d = 80$~nm with $l_r = 260$~nm. This finding suggests that surface scattering contributes at least partially to the relaxation. At small $l$, both curves meet indicating a diameter independent transmission probability $T$. Comparing the observed mean free path to theoretical predictions, we used the conventional Fr{\"o}hlich
coupling within the two-band Kane model~\cite{white1981, bastard1981}. This calculation results in a relaxation rate of $l_r = 230$~nm in excellent agreement with the experiments, see methods for details. 
The non-parabolicity of the conduction band is important for the relaxation lenght: neglecting it results in five times larger $l_r$ which is not consistent with the experiments.

In conclusion, we demonstrated transport in the ballistic limit with three-terminal hot carrier nanowire devices. The structures were made with 1D epitaxy enabling flexible band-gap engineering with smooth and sharp interfaces without doping. We determined the mean-free path and barrier transmission probability directly from the experiments at room temperature. For shortest devices, most of the electrons ($70\ \%$) are ballistic. We also found the mean-free path to depend on the nanowire diameter. These findings are consistent with the theoretical estimates. 

\section{Acknowledgements}
We thank the Knut and Alice Wallenberg (KAW) foundation via Project No. 2016.0089, Swedish Research Council (Dnr 2019-04111) and NanoLund for financial support.

\section{Methods}

\subsection{Nanowire heterostructure growth} 
Various growth stages in chemical beam epitaxy (CBE) system were systematically followed to grow the InAs/InP based nanowire heterostructures using Au aerosol particles (diameter $\sim$ 60 nm) deposited on InAs(111)B substrates. The precursors used were trimethylindium (TMIn), precracked tertiarybutylarsine (TBAs) and precracked tertiarybutylphosphine (TBP) for In, As and P, respectively. The first growth step was the removal of surface oxides from substrate which was carried out through annealing of the substrate at 540 °C under As pressure. Then the substrate temperature was lowered and stabilized to 420 °C. Pressures for different sources in the lines were controlled and measured prior to their entrance in growth chamber. Onset of growth is defined through flowing In into the chamber with fixed flow of TBAs. First, a InAs stem was grown using 0.1 mbar TMIn and 1.5 mbar TBAs. Subsequently, substrate temperature was further lowered and stabilized at 400 °C. Next, the operational segments of the nanowire heterostructure were grown at this stabilized temperature. The segments were grown by controlling the length with growth time in the following order: InAs emitter, $\mathrm{InAs}_{1–x}\mathrm{P}_x$, based graded barrier, InAs base, InP based filter barrier and InAs collector. The composition between different segments was set by changing and switching the partial pressures of the precursors. The compositional grading was grown using a constant TMIn pressure of 0.1 mbar, constant TBP pressure of 1 mbar and varying the TBAs pressure from 1.5 mbar to the withdrawal from the chamber. A number of growth runs were performed to obtain devices with different base lenght $l$.

\subsection{Nanowire device processing} 
After the growth, the three terminals were made to form the hot electron nanowire devices. The processing steps are as follows: The grown nanowires were transferred to a cleaned Si substrate capped with a 100 nm thick thermal oxide. The substrate was additionally pre-processed to have the lithographically defined alignment marks and metal pads for electrical contacts. After nanowire transfer, their location with respect to the alignement marks was determined and an electron beam lithography was performed to contact the wires. Just before metallization, ammonium polysulfide ((NH$_4$)$_2$S$_x$) solution based process was employed to remove the native oxide from the exposed InAs leads as well as to passivate the surface to avoid reoxidation~\cite{suyatin2007}. Ohmic metal contacts (25 nm Ni and 125 nm Au) to the InAs leads were deposited by thermal evaporation followed by lift-off. 

\subsection{Energy relaxation with the Kane model}
{\bf Energy dispersion:} To consider the energy relaxation in the base electrode, we use the so-called Kane model~\cite{white1981, bastard1981} implying the dispersion
\begin{equation}
E(k)=
E_c+\frac{-\Delta}{2}+\frac{\Delta}{2}\sqrt{1+2\frac{\hbar^2k^2}{m^*\Delta}}
=E_c+\frac{\hbar^2k^2}{2\, m(E)},
\label{eq:Kane}
\end{equation}
where
$E_c$ is the conduction band edge, $m(E)=m^*(1+(E-E_c)/\Delta))$ the energy dependent effective mass, $m^*$ the effective mass at the conduction band edge and  $\Delta$ the energy gap. As we consider energies far from the conduction band edge, we take non-parabolicity into account in Eq.~(\ref{eq:Kane}). This provides the energy dependence of the quasi-momentum and velocity as
\begin{equation}
\left\{
\begin{array}{ccl}
  k(E) &=& \frac{1}{\hbar} \sqrt{2m^*(E-E_c)\left(1+\frac{E-E_c}{\Delta}\right)} \vspace{4pt} \\
 v(E) &=&
\frac{\hbar k(E)}{m^*}\left(1+2\frac{E-E_c}{\Delta}\right)^{-1}.
\end{array}
\right.
\label{eq:kvKane}
\end{equation}
We use $E_c=0$, $\Delta=0.354$~eV and $m^*=0.023\, m_e$ for InAs as well as
$E_c=\Phi_B=0.5$~eV, $\Delta=1.344$~eV and $m^*=0.08\, m_e$ for InP barrier. 
For the kinetic energy of $E = 0.5$~eV in the InAs base contact, Eq.~(\ref{eq:kvKane}) yields a velocity of $v(E) = 1.12\times 10^6\ \textrm{m}/\textrm{s}$. For comparison, using the standard parabolic bandstructure in the conduction band, we obtain $v(E) = \sqrt{2E/m^*}=2.77\times 10^6\ \textrm{m}/\textrm{s}$, which demonstrates the huge impact of non-parabolicity.

{\bf Phonon scattering:} The scattering rate for the spontaneous emission of polar optical phonons via the
Fr{\"o}hlich interaction is
\begin{equation}
W^{\text{polar LO}}_{{\bf k}\to {\bf k}'}=\frac{2\pi}{\hbar}
\frac{\hbar^2A_\textrm{Fr}}{V}
\frac{
\delta\left(E({\bf k}')-E({\bf k})+\hbar\omega_{LO}\right)}{|{\bf k}-{\bf k}'|^2}
\label{EqWoptPhon}
\end{equation}
with the material constant
\[
A_\textrm{Fr}=\frac{\omega_{LO}e^2}{2\epsilon_0\hbar}
\left(\frac{1}{\epsilon(\infty)}-
\frac{1}{\epsilon_r}\right)=9.58\times 10^{18}\frac{\textrm{m}}{\textrm{s}^2}.
\]
Here we use the optical phonon energy
$\hbar\omega_{LO}=30$ meV, the static dielectric constant
$\epsilon_r=15.15$, as well as its high frequency value 
$\epsilon(\infty)=12.3$ which are common values for bulk InAs.
For an isotropic band structure $E({\bf k})=E(k)$ with the Kane model addressed above, we obtain the total spontaneous LO phonon emission rate for an electron in the initial state ${\bf k}$ as
\begin{equation}
\begin{array}{l}
\displaystyle \frac{1}{\tau_\mathrm{LO\, emis.}} =
  \frac{V}{(2\pi)^3}\int {\rm d}^3 k'\, W^{\text{polar LO}}_{{\bf k}\to {\bf k}'} = \vspace{6pt} \\
   \frac{A_\mathrm{Fr}m^*}{2\pi\hbar}
   \frac{1+2(E_0-\hbar\omega_{LO})/\Delta}{k}
     \log\left|\frac{k+k_f}{k-k_f}\right|\Theta(E_0-\hbar\omega_{LO}),
 \end{array}
 \end{equation}
where the  momenta $k,k_f$ depend on the initial energy $E_0$ as
\begin{equation}
\left\{
\begin{array}{ccl}
     \hbar k &=& \sqrt{2m^*E_0+2m^*\frac{E_0^2}{\Delta}} \vspace{6pt} \\
     \hbar k_f &=& \sqrt{2m^*(E_0-\hbar\omega_{LO})+2m^*\frac{(E_0-\hbar\omega_{LO})^2}{\Delta}}.
 \end{array}
\right.
\end{equation}
 This provides the scattering time $\tau_\mathrm{LO\, emis.} = 207$ fs for the initial energy $E_0=0.5$ eV at room temperature. Together with the velocity of $1.12\times 10^6$ m/s addressed above, we obtain a relaxation length $l_r=230$ nm. Here we disregard the stimulated processes as the energy balance for stimulated emission and absorption approximately compensate each other. Given the fact that the scattering mostly results only in small changes in the Bloch vector, see Eq.~(\ref{EqWoptPhon}), the velocity is hardly changed and the net energy loss by spontaneous emission dominates the odds of the injected electron to overcome the barrier.
We note, that the use of a parabolic band provides more than twice the scattering time, due to the reduced density of states. In combination with the increased velocity, this provides relaxation lengths which are about five times larger and not consistent with the experimental findings.

{\bf Impurity scattering:} We use the common screened impurity potential
\begin{equation}
V({\bf r})=\frac{ e^2 \mathrm{e}^{-\lambda r}}{4\pi \varepsilon_r \epsilon_0 r}=
\frac{1}{V}
\sum_{\bf q} \frac{e^2}{ \varepsilon_r\epsilon_0(q^2+\lambda^2)} \mathrm{e}^{{\rm i}{\bf q}\cdot{\bf r}}
\end{equation}
with the Debye screening length $\lambda^2=ne^2/\varepsilon_r\epsilon_0k_BT$, where $n$ is the electron density and we apply  the temperature $T=300$ K.
Fermi's golden rule provides for $N_i$ independent impurities
\begin{equation}
\begin{array}{l}
\frac{1}{\tau_\mathrm{imp}}
= N_i\frac{2\pi}{\hbar}\frac{V}{(2\pi)^3}\int {\rm d}^3 k' \vspace{6pt} \\
\left|\frac{e^2}{V \varepsilon_r\epsilon_0(|{\bf k}-{\bf k'}|^2+\lambda^2)}\right|^2\delta\big(E({\bf k})-E({\bf k}')\big) \vspace{6pt} \\
= \frac{N_i}{V}\frac{e^4}{4\pi \hbar^2\varepsilon_r^2\epsilon_0^2}  \frac{1}{v(k)}
\left(\frac{1}{\lambda^2}-\frac{1}{4k^2+\lambda^2}\right).
 \end{array}
\label{EqImpScatt}
\end{equation}
Note that
$\frac{N_i}{V\lambda^2}$ does not depend on the doping in case of charge
neutrality $n=N_i/V$, which is assumed here. For $k$ corresponing to the initial energy of 0.5 eV, we obtain with the Kane model $\tau_\mathrm{imp}=198$ fs for a
doping density of $10^{17}/\textrm{cm}^3$. This is very similar to the spontaneous phonon emission time calculated above. However most of the scattering is forward and the common momentum relaxation time 11.5 ps is significantly longer, so that impurity scattering should not limit the transmission over the barrier. This result is obtained by inserting the factor $(1-\cos[\angle({\bf k},{\bf k'})])$ into the integral in Eq.~(\ref{EqImpScatt}). 

Electron-electron scattering matrix elements are structurally
similar to the impurity scattering, but due to exchange effects
one expects a reduction of scattering rate by a factor of four
\cite{Dur1996,Callebaut2004}. Thus we neglect these here. An open issue are Auger processes, which we did not consider.

{\bf The effect of crystal structure:} Above we applied the parameter values for bulk InAs, which are well established. A major issue is the fact, that the experimental nanowires are of wurtzite type.  Band structure calculations in Ref. \citealp{fariajunior2016} provide a linear energy-$k$ relation with $v_D\approx 0.9 \times 10^6 $ m/s in the range of interest, which is quite comparable to the zincblende case: Reference~\citealp{moller2011} reports $\omega_{LO}=2\pi c \cdot 239$/cm, which agrees well with the zincblende value of $\omega_{LO}$. Furthermore  $\omega_{TO}=2\pi c \cdot 214$/cm, provides a ratio $\omega_{LO}/\omega_{TO}=1.12$, which agrees well with the Lyddane-Sachs-Teller relation $\sqrt{\varepsilon_r/\varepsilon(\infty)}=1.11$ for zincblende. A slightly smaller value $\omega_{LO}/\omega_{TO}=236/219=1.08$ is reported in Ref.~\citealp{Dabhi2015}. In summary, the scattering rate should not differ significantly between  wurtzite and zincblende material.

Finally, it is worth to mention that the splitting between the first and second conduction band in wurtzite InAs nanowires was recently measured to equal to 590~meV \cite{Pournia2020}, which might provide complications not considered here.

\subsection{Transmission probability}
\begin{figure}[t]
    \centering
    \includegraphics[width=0.3\textwidth]{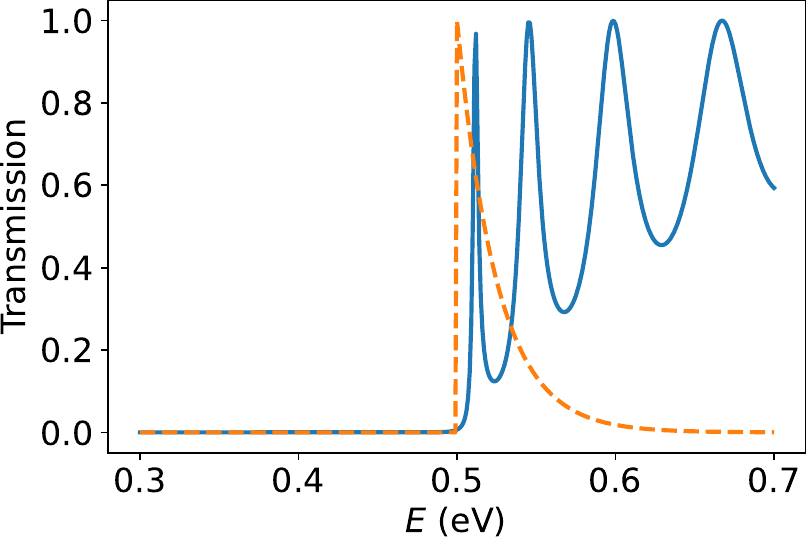}
    \caption{Transmission (full line) through the InP barrier with a thickness of $a=20$ nm using the Kane model together with the thermal distribution (dashed line) at room temperature, which is used for averaging.}
    \label{fig:transmission}
\end{figure}

The transmission through the barrier of thickness $d$ can be evaluated in a standard way~\cite{davies2006}, resulting in the transmission probability
\begin{equation}
T(E)=\left[1+\left|\frac{m_wk_b}{m_bk_w}- \frac{m_bk_w}{m_wk_b} \right|^2\left| \sin(k_b a) \right|^2  \right]^{-1},
\end{equation}
where the $m_{w/b}$ and  $k_{w/b}$ are the functions $m(E)$ and $k(E)$ given in Eqs.~(\ref{eq:Kane}) and (\ref{eq:kvKane}) with InAs and InP parameters respectively. Note, that the expression holds both for $E>\Phi_B$ and for $E<\Phi_B$ with $k_b$ being imaginary. The result is shown in Fig.~\ref{fig:transmission} with thermal Bolzmann-distribution $\propto \exp(-E/k_BT)$ for $E > \Phi_B$.

\end{document}